\documentstyle[11pt,paspconf,epsf]{article}

\def\etal{{et al.\ }}
\def\Lya{Ly$\alpha\ $}
\def\HI{\hbox{H~$\scriptstyle\rm I\ $}}
\def\msun{{\rm\,M_\odot}}

\def\ub{U-B}
\def\bv{B-V}
\def\vi{V-I}
\def\bi{B-I}
\def\spose#1{\hbox to 0pt{#1\hss}}
\def\lta{\mathrel{\spose{\lower 3pt\hbox{$\mathchar"218$}}
     \raise 2.0pt\hbox{$\mathchar"13C$}}}
\def\gta{\mathrel{\spose{\lower 3pt\hbox{$\mathchar"218$}}
     \raise 2.0pt\hbox{$\mathchar"13E$}}}

\def\iunits{\,{\rm ergs\,cm^{-2}\,s^{-1}\,Hz^{-1}\,sr^{-1}}}
\def\sfrd{\,{\rm M_\odot\,yr^{-1}\,Mpc^{-3}}}
\def\mdden{\,{\rm M_\odot\,Mpc^{-3}}}

\begin{document}

\title{The Hubble Deep Field and the Early Evolution of Galaxies}
\author{Piero Madau}
\affil{Space Telescope Science Institute, 3700 San Martin Drive, Baltimore,
MD 21218}

\begin{abstract}

I review some recent progress made in our understanding of galaxy evolution and
the cosmic history of star formation. The {\it Hubble Deep Field} (HDF)
imaging survey has achieved the sensitivity to capture the bulk of the
extragalactic background light from discrete sources. No evidence is
found in the optical number-magnitude relation down to $AB=29$
mag for a large amount of star formation at high redshifts. A census of
the ultraviolet and blue ``dropouts'', which requires the inclusion of
the effects of intergalactic attenuation on the colors of cosmologically
distant galaxies, appears to confirm this basic conclusion. The emission 
history of the universe 
at ultraviolet, optical, and near-infrared wavelengths can be modeled from the 
present epoch to $z\approx 4$ by tracing the evolution with cosmic time of the 
galaxy luminosity density, as determined from several deep spectroscopic 
samples and the HDF. The global spectrophotometric properties of
field galaxies are well fitted by a simple stellar evolution model, defined
by a time-dependent star formation rate (SFR) per unit comoving volume and a
universal initial mass function which is relatively rich in massive
stars. The SFR density is found to rise sharply, by about an order 
of magnitude, from a redshift of zero to a peak value  at $z\approx 1.5$ 
in the range 0.12--0.17 $\sfrd$, to fall again by a factor of 2 (4) out of 
a redshift of 3 (4). Since
only 10\% of the current stellar content of galaxies is produced at $z>2.5$, a
rather low cosmic metallicity is predicted at these early times, in good
agreement with the observed enrichment history of the damped Lyman-$\alpha$
systems. The biggest uncertainty is represented by the poorly constrained
amount of starlight that was absorbed by dust and reradiated in the IR
at early epochs.  A ``monolithic collapse'' scenario, where half of the
present-day stars formed at $z>2.5$ and were shrouded by dust, can be made
consistent with the global history of light, but appears to overpredict the 
metal mass density at high redshifts.

\end{abstract}


\section{Introduction}

Much observing time has been devoted in the past few years to the problem of
the detection of galaxies at high redshifts, as it was
anticipated that any knowledge of their early luminosity and color evolution
would set important constraints on the history of structure and star formation
in the universe. While it has become now clear that blank-sky surveys for
strong \Lya-emitting primeval galaxies are not particularly efficient 
(Djorgovski \& Thompson 1993), the method of obtaining multicolor broadband
observations  of the emitter's rest-frame UV and optical stellar continuum 
has been successfully applied to select galaxies at cosmological distances in
ground-based surveys (Steidel \& Hamilton 1992; Steidel \etal 1996a; see
review by Pettini in this proceedings) and in the {\it Hubble Deep Field} 
(HDF) (Madau \etal 1996, hereafter
M96; Steidel \etal 1996b; Lowenthal \etal 1997). Together with the tremendous
progress in our understanding of faint galaxy data at $z\lta 1$ made possible
by the recent completion of several comprehensive ground-based spectroscopic
surveys (Lilly \etal 1995; Ellis \etal 1996; Cowie \etal 1996), the
identification of star-forming galaxies at $2\lta z\lta 4$ has provided new
clues to some key questions of galaxy formation and evolution studies: Is there
a characteristic epoch of star and metal formation in galaxies?  What fraction
of the luminous baryons observed today were already locked into galaxies at
early epochs? Are high-$z$ galaxies obscured by dust? Do spheroids form early
and rapidly? Is there a ``global'' IMF? Through the systematic study of
galaxies at increasing cosmological lookback times it has become possible to
reconstruct the history of stellar birthrate {\it directly}, as opposite to the
``classical'' method where one studies the resolved stellar populations of the
Milky Way and nearby galaxies and infers their evolutionary history from fossil
records -- well-known examples of this more traditional approach are nuclear
cosmochronology, the color-magnitude diagram of globular clusters, the cooling
sequence of white dwarfs (see Renzini 1993 and references therein). 

In this talk I will review the broad picture that has recently emerged from the
``direct'' method, focusing on the emission properties at ultraviolet, optical,
and near-IR wavelengths of the galaxy population {\it as a whole}. I will show
how the combination of HST deep imaging and ground-based spectroscopy offers 
now an exciting first glimpse to the history of the conversion of neutral gas
into stars in field galaxies. In the following, all magnitudes will be given 
in the AB system,
and a flat cosmology with $q_0=0.5$ and $H_0=50\,$km s$^{-1}$ Mpc$^{-1}$ will
be adopted. 

\section{The Hubble Deep Field}

As the best view to date of the optical sky at faint flux levels, the
HDF imaging survey has rapidly become a key testing ground
for models of galaxy evolution. The field, an undistinguished portion of the
northen sky at high galactic latitudes, was imaged for approximately 150 orbits
from 18 to 30 December 1995 with the Wide Field Planetary Camera onboard the
{\it Hubble Space Telescope}. With its depth -- reaching 5-$\sigma$ limiting AB
magnitudes of roughly 27.7, 28.6, 29.0, and 28.4 in $U, B, V,$ and $I$
\footnote{This is roughly three magnitudes fainter than the deepest ground
based images in the red bands, two magnitudes deeper in the blue, and one
magnitude deeper in the ultraviolet.}~(Williams \etal 1996) -- and four-filter
strategy in order to detect Lyman-break galaxies at various redshifts, the HDF
offers the opportunity to study the galaxy population in unprecedented detail. 

\begin{figure}
\plotfiddle{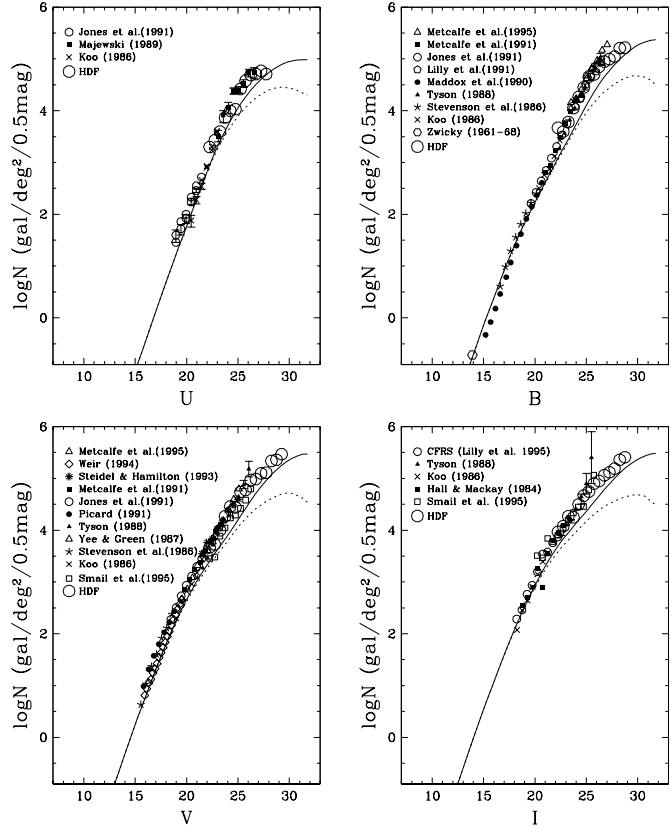}{4.0in}{0}{50}{50}{-150}{-40}
\caption{Differential galaxy number counts per square degree per half magnitude
interval as a function of AB apparent magnitude. The sources of the data points
are indicated in each panel (see Pozzetti \etal 1997 for the complete
reference list). Note the decrease of the logarithmic slope $d\log
N/dm$ at faint magnitudes. The curves show the well known deficiency of faint
galaxies which characterizes no-evolution models in a $\Omega=0$ ({\it solid
line}) and $\Omega=1$ ({\it dotted line}) universe. 
}
\label{fig1}
\end{figure}

\subsection {Galaxy Counts and the EBL} 

There are about 3,000 galaxies in the HDF, corresponding to $2\times 10^6$
deg$^{-2}$ to $V\approx 29$ mag. The galaxy counts as a function of AB
isophotal magnitude are shown in Figure 1 in the F300W, F450W, F606W, and F814W
bandpasses (the number corresponds to the central wavelength in nm) for all
galaxies with signal-to-noise ratio $S/N>3$ within the band (Williams \etal
1996). A compilation of existing ground-based data is also shown, together with
the predictions of no-evolution models, i.e. models in which the absolute
brightness, volume density, and spectra of galaxies do not change with time. 
The HDF counts are plotted down to the 80\% completeness limit and are found to
agree reasonably well with previous surveys, to within $20\%$ in the magnitude
range $22<AB<26$.  In all four bands, the slope $\alpha$ of the differential
galaxy counts, $\log N(m)=\alpha m$, flattens at faint magnitudes, e.g., from
$\alpha=0.45$ in the interval $21<B<25$ to $\alpha=0.17$ for $25<B<29$. This
feature cannot be due to the reddening of distant sources as their Lyman break
gets redshifted into the blue passband, since the fraction of Lyman-break
galaxies at $B\sim 25$ is only of order 10\% (cf. Guhathakurta \etal 1990).
Moreover, an absorption-induced loss of sources could not explain the similar
flattening of the number-magnitude relation observed in the $V$ and $I$ bands.
Rather, the change of slope suggests a decline in the surface density of
luminous galaxies beyond $z\sim 1.5$. 

The contribution of known galaxies to the extragalactic background light (EBL)
-- an indicator of the total optical luminosity of the universe -- can be
calculated directly by integrating the emitted flux times the differential
galaxy number counts down to the detection threshold. The leveling off of the
counts is clearly seen in Figure 2, where the function
$i_\nu=10^{-0.4(m+48.6)}\times N(m)$ is plotted against apparent magnitude in
all bands (Pozzetti \etal 1997). While counts having a logarithmic slope of
$\alpha\ge0.40$ continue to add to the EBL at the faintest magnitudes, it
appears that the HDF survey has achieved the sensitivity to capture the bulk of
the extragalactic light from discrete sources (an extrapolation of the observed
counts to brighter and/or fainter magnitudes would typically increase the
sky brightness by less than 20\%). To $AB=29$, the sky brightness from
resolved galaxies in the $I$-band is $\approx 2\times 10^{-20}\iunits$,
increasing roughly as $\lambda^2$ from 2000 to 8000 \AA.
The  flattening of the number counts has
the interesting consequences that the galaxies that produce $\sim 60\%$ of the
blue EBL have $B<24.5$. They are then bright enough to be identified in
spectroscopic surveys, and are indeed known to have median redshift $\langle
z\rangle=0.6$ (Lilly \etal 1995). The quite general conclusion is that there 
is no evidence in the number-magnitude relation down to very faint flux 
levels for a large amount of star formation at high redshift. Note that
these considerations do not constrain the {\it rate} of starbirth at early
epochs, only the total (integrated over cosmic time) amount of stars -- hence
background light -- being produced, and {\it neglect the effect of dust 
reddening}. 

\begin{figure}
\plotfiddle{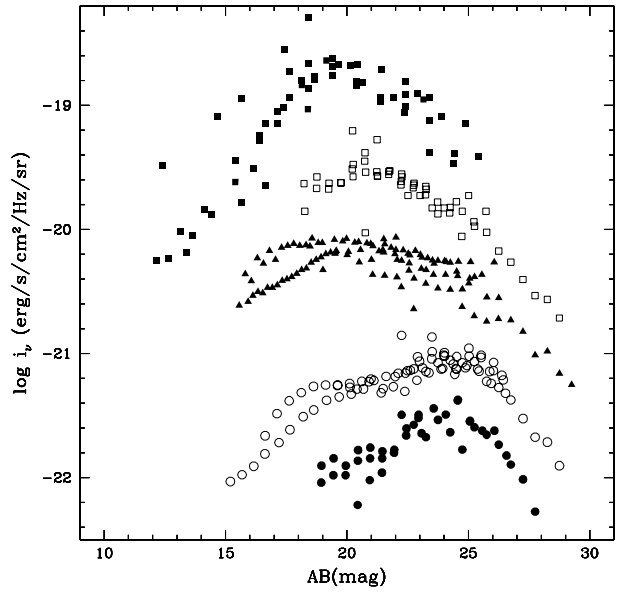}{3.5in}{0}{50}{50}{-150}{-80}
\caption{The contribution of known galaxies to the extragalactic background 
light per magnitude bin as a function of $U$ ({\it
filled circles}), $B$ ({\it open circles}), $V$ ({\it filled triangles}), $I$
({\it open squares}) and $K$ ({\it filled squares}) magnitudes. For clarity, 
the $B$, $V$, $I$ and $K$ values have been multiplied by a factor of 2, 8,
20, and 40, respectively.
}
\label{fig2}
\end{figure}

\subsection{Intergalactic Absorption}

A more direct way to track galaxy evolution at early epochs is through a census
of the HDF ``dropouts''. 
At faint magnitudes, the interpretation and detailed modeling of the
observations require the self-consistent inclusion of the effect of
intergalactic attenuation on galaxy colors.
Absorption by intervening material has been known for quite some time to
distort our view of objects at cosmological distances. It has been realized
only recently, however, that the accumulated line-blanketing and Lyman-continuum
absorption from the \Lya forest clouds and Lyman-limit systems along the path 
to high redshifts can be efficiently used to identify galaxies
at $z\gta2$ (Madau 1995; Steidel \& Hamilton 1992). Although other spectral
features, such as the $4000\,$\AA\ and $912\,$\AA\ breaks which characterize the
integrated spectra of stellar populations, with the latter possibly enhanced by
self-absorption from interstellar gas within the galaxy itself, can and have
been used as tracers of redshifts, the model predictions of their
magnitude are sensitive to the unknown physical and evolutionary state of the
galaxy, i.e., its star formation history, age, and \HI distribution, and hence
are subject to substantial uncertainties.  By contrast, the ``reddening''
effect due to atomic processes in cosmological distributed QSO absorption
systems is ubiquitous and can be reliably taken into account. While
stochastic in nature, r.m.s. fluctuations away from the mean opacity are bound
to be modest in most situations, due to the broadband nature of the adopted
filter set. 

\begin{figure}
\plotfiddle{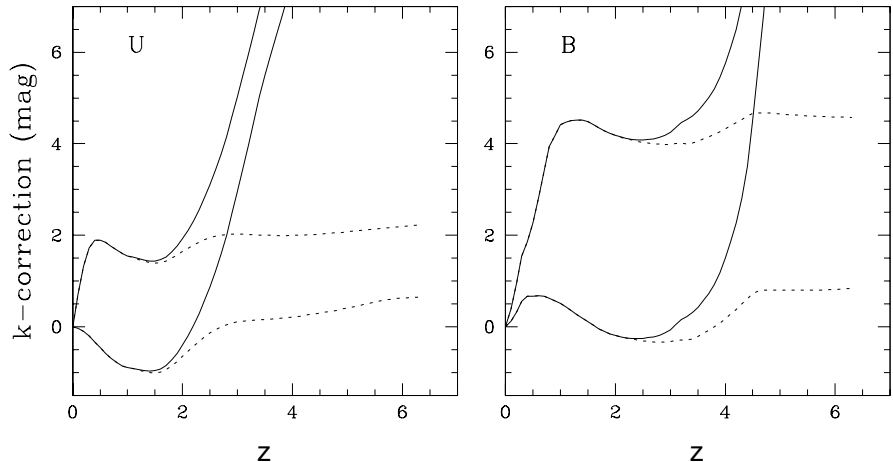}{3.5in}{0}{80}{80}{-250}{-315}
\caption{Average $k$-correction in the $U$ and $B$ bands for
elliptical and late-type spiral galaxies  as a function of redshift. {\it
Dotted lines:} unattenuated synthetic spectra. {\it Solid lines:} spectra
modified by intergalactic absorption.
}
\label{fig3}
\end{figure}

The magnitude and cosmological importance of
intergalactic absorption on galaxy spectra can be effectively illustrated by
including its effect in the standard $k$-correction term needed to translate
the galaxy magnitude at Earth into its rest-frame value, 
\begin{equation}
k(z)=-2.5\log\left[(1+z){L(\nu)\over L(\nu_{\rm obs})}\langle
e^{-\tau}\rangle \right], 
\end{equation}
where $L(\nu)$ is the specific power emitted by a source at redshift $z$,
$\nu_{\rm obs}=\nu/(1+z)$, $\langle e^{-\tau}\rangle$ is the cosmic
transmission averaged over all lines of sight, and we have assumed no intrinsic
luminosity evolution. Figure 3 shows the $k$-correction in the $U$ and
$B$ HDF bands as a function of redshift for synthetic spectra of galaxies
which well reproduce the colors
of present-day ellipticals and spirals. The effect at high redshift is huge.
Due to intergalactic attenuation alone, the $k$-correction in the F300W
bandpass increases by as much as 4 magnitudes between $z\approx 2$ and
$z\approx 3.5$, giving origin to a ``UV dropout''. In the F450W band, the
increase is barely noticeable at $z\approx3$, but becomes very large above
$z\approx 4$, thereby producing a ``blue dropout''. 

\begin{figure}
\plotfiddle{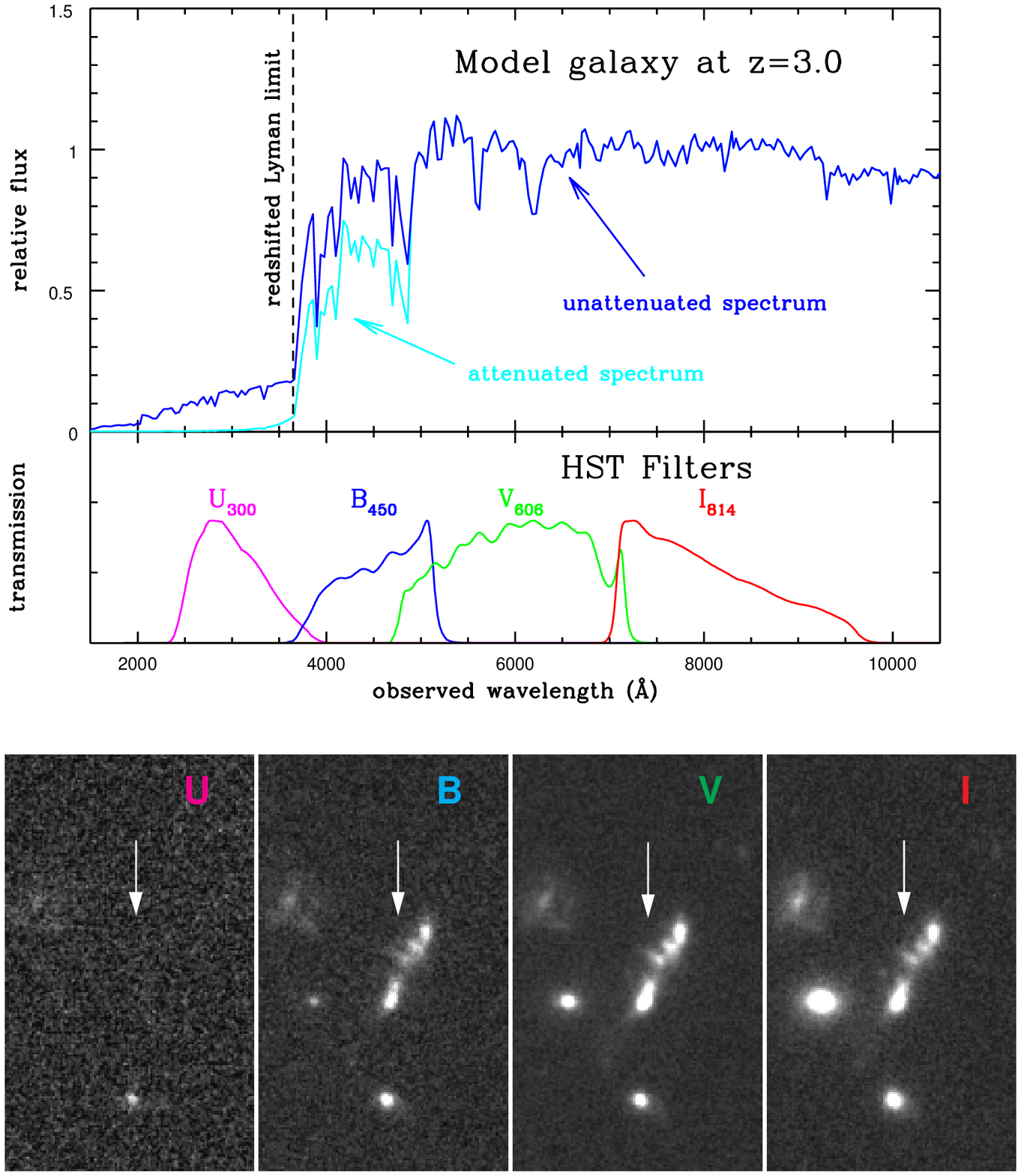}{5.7in}{0}{80}{80}{-250}{-50}
\caption{Synthetic spectrum of a star-forming galaxy at $z=3$, superimposed on
the throughput curves of the HDF filters. Such a galaxy would appear blue in its
optical colors but would be virtually undetected in the ultraviolet filter. At 
higher redshifts ($z\gta4$), a similar galaxy would vanish from the blue filter.
{\it Bottom panel}: $U, B, V,$ and $I$ images of an identified UV dropout at
$z=2.8$. (Courtesy of M. Dickinson.) 
}
\label{fig4}
\end{figure}

\subsection {Ultraviolet and Blue Dropouts} 

Ground-based observations have used color techniques which are sensitive to the
presence of a Lyman-continuum break superposed to an otherwise flat UV spectrum
to identify galaxies at $z\approx 3$ (Steidel \& Hamilton 1992; Steidel \etal 
1996a). 
New photometric criteria for robustly selecting Lyman-break galaxies have been
developed based on the HDF color system, providing what appear to be largely
uncontaminated samples of star-forming galaxies at high redshifts 
(M96). The HDF ultraviolet passband -- which is bluer than the standard
ground-based $U$ filter -- permits the identification of star-forming galaxies
in the interval $2<z<3.5$ (Figure 4). Galaxies in this redshift range
predominantly occupy the top left portion of the $\ub$ vs. $\bi$ color-color
diagram because of the attenuation by the intergalactic medium and intrinsic
extinction. Galaxies at lower redshift can have similar $U-B$ colors, but they
are typically either old or dusty, and are therefore red in $B-I$ as well. Of
order 100 (200) ultraviolet dropouts can be identified in the HDF which are
brighter than $B=27$ ($B=29$), approximately 25\% (20\%) of the total. To
date, about 25 among the brightest ones have spectroscopically 
confirmed redshift in the range
$2.0<z<3.4$. The color-selection region is illustrated in Figure 5. The $UBI$
criteria isolate objects that have relatively blue colors in the optical, but a
sharp drop into the UV. 
In analogous way, the blue passband allows the selection of candidate
star-forming galaxies in the interval $3.5<z<4.5$. Only
$\sim 20$ (60) $B$ dropouts down to $V=28$ ($V=30$) have been identified
in the $\bv$ vs. $\vi$ plane (M96). The brightest one has recently 
been confirmed through deep Keck
spectroscopy to be at $z=4.02$  (Dickinson 1997), consistent with the
photometric predictions. 

\begin{figure}
\plotfiddle{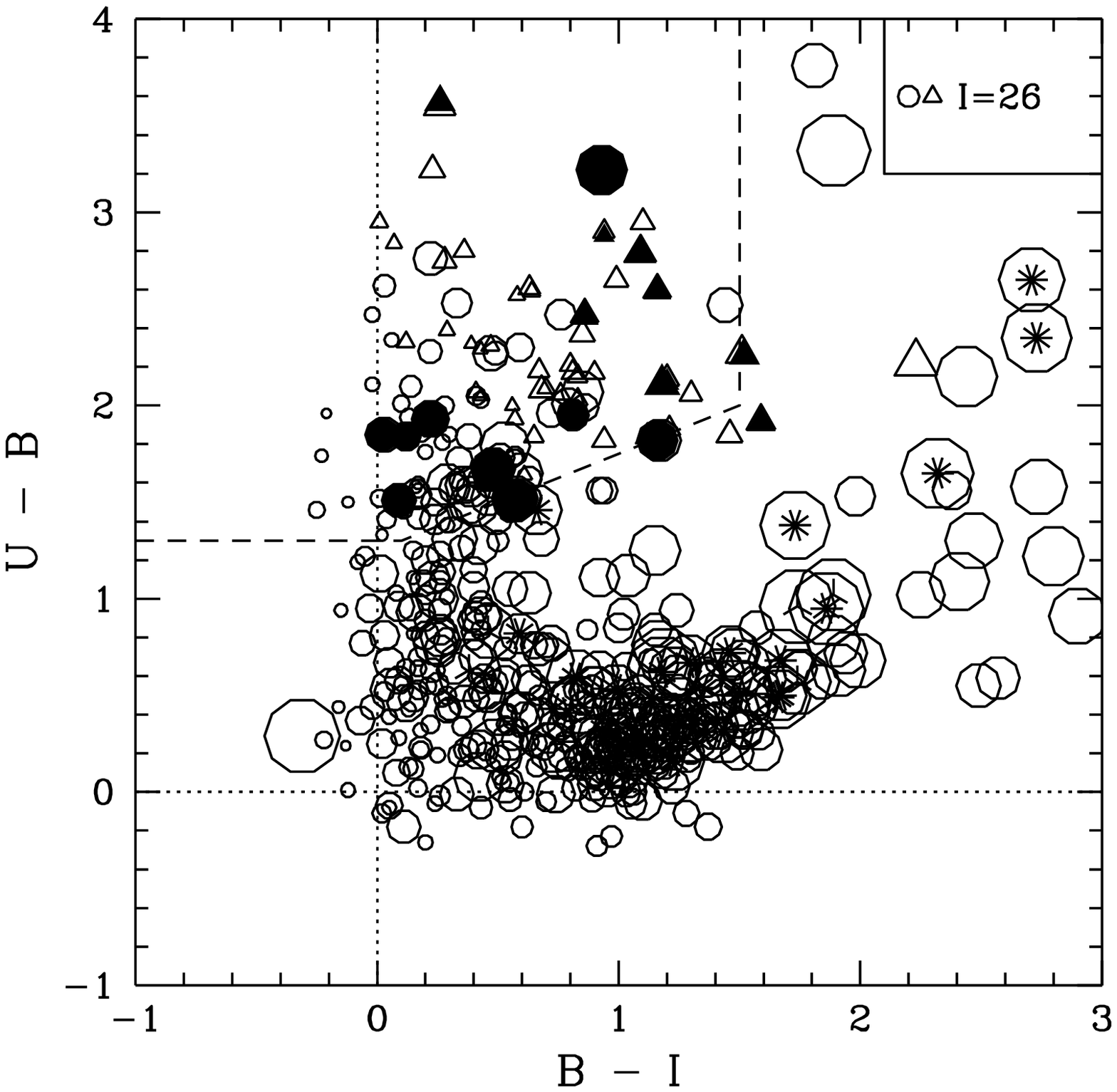}{3.5in}{0}{50}{50}{-150}{-80}
\caption{Color-color plot of galaxies in the HDF with $B<27$. Objects
undetected in $U$ (with $S/N<1$) are plotted as triangles at the 1$\sigma$
lower limits to their $U-B$ colors. Symbols size scales with the $I$ mag of the
object. The dashed lines outline the selection region within which we identify
candidate $2<z<3.5$ objects. Galaxies with spectroscopically confirmed
redshifts within this range (Steidel \etal 1996b;
Lowenthal \etal 1997) are marked as solid symbols. Galaxies having confirmed 
redshifts less than 2 (Cohen \etal 1996) are marked as asterisks. Note that 
no low-redshift interlopers have been found among the high-redshift sample. 
}
\label{fig5}
\end{figure}

The UV continuum emission from a galaxy with significant ongoing 
star formation is entirely dominated by late-O/early-B stars on the main 
sequence, which have masses $\gta 10\msun$ and lifetimes $t_{MS}\lta 
2\times 10^7\,$yr. After an initial transient phase where the UV flux
rises rapidly and the turnoff mass drops below $10\msun$, a steady 
state is reached where the measured luminosity becomes proportional to the 
instantaneous SFR and independent of the past star formation history
(see Madau, Pozzetti, \& Dickinson 1997).
Figure 6 depicts the present-epoch ``SFR function'' -- which describes the
number of star-forming galaxies as a function of their ongoing SFR --, 
together with the derived distribution of stellar birthrates for the HDF
ultraviolet and blue dropouts. The comparison shows the sign of a 
significant density evolution -- a rapid increase in $\phi_*$.

\begin{figure}
\plotfiddle{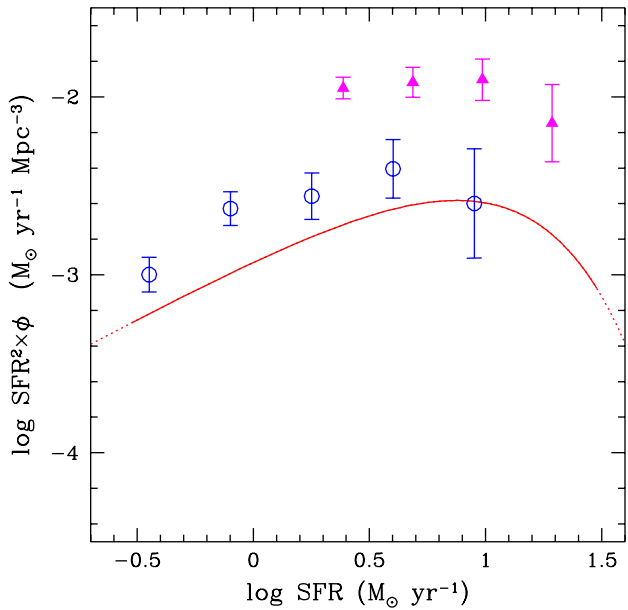}{3.5in}{0}{50}{50}{-150}{-80}
\caption{Distribution of star formation rates at different redshifts.
{\it Filled triangles}: HDF ultraviolet dropouts at $\langle
z\rangle=2.75$. {\it Empty circles}: HDF blue droputs at $\langle
z\rangle=4$. The curve represents the best-fit Schechter function from
the Gallego \etal (1995) H$\alpha$ survey at $z=0$.
}
\label{fig6}
\end{figure}

\section{Evolution of the Galaxy Luminosity Density}

The integrated light radiated per unit volume from the entire galaxy population
is an average over cosmic time of the stochastic, possibly short-lived star
formation episodes of individual galaxies, and should follow a relatively simple
dependence on redshift. In the UV -- where it is proportional to the global 
star formation rate (SFR) -- its evolution should provide information, e.g., on
the mechanisms which may prevent the gas within virialized dark matter halos to
radiatively cool and turn into stars at early times, or on the epoch when
galaxies exhausted their reservoirs of cold gas. From a comparison between
different wavebands it should be possible to set constraints on the average 
initial mass function (IMF) and dust content of galaxies. 

\begin{figure}
\plotfiddle{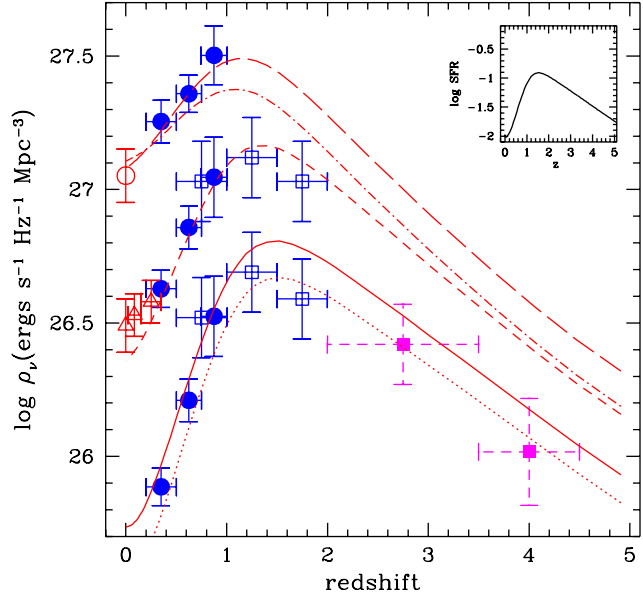}{3.5in}{0}{50}{50}{-150}{-80}
\caption{Evolution of the luminosity density at rest-frame wavelenghts of 0.15
({\it dotted line}), 0.28 ({\it solid line}), 0.44 ({\it short-dashed line}),
1.0 ({\it long-dashed line}), and 2.2 ({\it dot-dashed line}) \micron. The data
points with error bars are taken from Lilly \etal (1996) ({\it filled dots}),
Connolly \etal (1997) ({\it empty squares}), M96 and Madau
(1997) ({\it filled squares}), Ellis \etal (1996) ({\it empty triangles}), and
Gardner \etal (1997) ({\it empty dot}). The inset in the upper-right corner of
the plot shows the SFR density ($\sfrd$) versus redshift which was used as input
to the population synthesis code. The model assumes a Salpeter IMF, SMC-type
dust in a foreground screen, and a universal $E(B-V)=0.06$.
}
\label{fig7}
\end{figure}

The comoving luminosity density, $\rho_\nu(z)$, from the present epoch to
$z\approx 4$ is shown in Figure 7 in five broad passbands centered around 0.15,
0.28, 0.44, 1.0, and 2.2 \micron.  The data are taken from the $K$-selected
wide-field redshift survey of Gardner \etal (1997), the $I$-selected  CFRS
(Lilly \etal 1996) and $B$-selected Autofib (Ellis \etal 1996) surveys, the
photometric redshift catalog for the HDF of Connolly \etal (1997) -- which take
advantage of deep infrared observations by Dickinson \etal (1997) -- and the
color-selected UV and blue ``dropouts'' of M96 (see also Madau 1997). They 
have all
been corrected for incompleteness by integrating over the best-fit Schechter
function in each redshift bin, 
\begin{equation}
\rho_\nu(z)=\int_0^\infty L\phi(L,z)dL=\Gamma(2+\alpha)\phi_*L_*. \label{eq:ld}
\end{equation}
As from the Connolly \etal (1997) and M96 data sets it is not possible to
reliably determine the faint end slope of the luminosity function, a value of
$\alpha=-1.3$ has been assumed at each redshift interval for comparison with
the CFRS sample (Lilly \etal 1995). The error bars are typically less than 0.2
in the log, and reflect the uncertainties present in these corrections and, in
the HDF $z>2$ sample, in the volume normalization and color-selection region.
In the $K$-band, the determination by Gardner \etal (1997) agrees to within
30\% with Cowie \etal (1996), and we have assigned an error of 0.1 in the log
to the estimate of the local luminosity density at 2.2 \micron. 
Despite the obvious caveats due to the likely incompleteness in the data sets,
different selection criteria, and existence of systematic uncertainties in the
photometric redshift technique, the spectroscopic, photometric, and Lyman-break
galaxy samples appear to provide a remarkably consistent picture of the
emission history of field galaxies. The UV luminosity density rises sharply, by
about an order of magnitude, from a redshift of zero to a peak at $z\approx
1.5$, to fall again by a factor of 2 (4) out to a redshift of 3 (4) (M96; Lilly
\etal 1996; Connolly \etal 1997). This points to a rapid drop in the
volume-averaged SFR in the last 8--10 Gyr, and to a redshift range $1\lta z\lta
2$ in which the bulk of the stellar population was assembled. The decline in
brightness at late epochs is shallower at longer wavelengths, as galaxies
becomes redder with cosmic time, on the average. 

\section{Stellar Population Synthesis Modeling}

Stellar population synthesis has become a standard technique to study the
spectrophotometric properties of galaxies. In the following, I will make 
extensive use of
the latest version of Bruzual \& Charlot (1993) isochrone synthesis code,
optimized with an updated library of stellar spectra (Bruzual \& Charlot 1997),
to predict the time change of the spectral energy distribution of a stellar
population. The uncertanties linked to the underlying stellar evolution
prescriptions and the lack of accurate flux libraries do not typically exceed
35\% (Charlot, Worthey, \& Bressan 1966). I will consider three
possibilities for the IMF, $\phi(m)\propto m^{-1-x}$: a Salpeter (1955)
function ($x=1.35$), a Scalo (1986) function,  which is flatter for low-mass
stars and significantly less rich in massive stars than Salpeter, and an
intermediate case with $x=1.7$. In all models the metallicity is fixed to 
{\it solar} values and the IMF is truncated at 0.1 and 125 $\msun$. 

An interesting question now arises as to whether a simple stellar evolution
model, defined by a time-dependent SFR per unit volume and a constant IMF, may
reproduce the global UV, optical, and near-IR photometric properties of the
universe as given in Figure 7. In a stellar system with arbitrary star formation
rate, the luminosity density at time $t$ is given by the convolution integral 
\begin{equation}
\rho_\nu(t)=\int^t_0 L_\nu(\tau)\times {\rm SFR}(t-\tau)d\tau, 
\label{eq:rho} 
\end{equation}
where $L_\nu(\tau)$ is the specific luminosity radiated per unit initial mass
by a generation of stars with age $\tau$. In the instantaneous recycling
approximation (Tinsley 1980), the total stellar mass density produced at time
$t$ is 
\begin{equation}
\rho_s(t)=(1-R)\int_0^t {\rm SFR}(t)dt, 
\end{equation}
where $R$ is the mass fraction of a generation of stars that is returned to the
interstellar medium, $R\approx 0.3, 0.15,$ and 0.2 for a Salpeter, $x=1.7$, and
Scalo IMF, respectively.

In computing the time 
evolution of the spectrophotometric properties of a stellar population in
comoving volumes large enough to be representative of the universe as a whole,
our first task is to relate the observed UV emission to a mean star formation 
rate. This is done by assuming a universal IMF and then fitting a 
smooth function to
the UV continuum emissivity at various redshifts. By construction, all models
will therefore produce, to within the errors, the right amount of 
ultraviolet light.  Bruzual and Charlot's synthesis code can then be used to 
predict the cosmic emission history at long wavelenghts. It is fair to
point out some of the limitations of this approach at the outset. (1)
It focuses on the emission properties of ``normal'', optically-selected 
field galaxies which are only moderately affected by dust -- a typical spiral
emits 30\% of its energy in the far-infrared region (Soifer \& Neugebauer
1991) --. Starlight which is completely blocked from view even in the near-IR
by a large optical depth in dust is not recorded by this technique, and the
associated baryonic mass and metals missed from our census. The contribution of
infrared-selected dusty starbursts to the integrated stellar mass density
cannot be large, however, for otherwise the current limits to the energy
density of the mid- and far-infrared background would be violated (Puget \etal
1996; Kashlinsky, Mather, \& Odenwald 1996; Fall, Charlot, \& Pei 1996;
Guiderdoni \etal 1997).
Locally, infrared luminous galaxies are known to produce only a small fraction
of the IR luminosity of the universe (Soifer \& Neugebauer 1991). (2) While
the method bypasses the ambiguities associated with the study of
morphologically-distint samples whose physical significance remains unclear,
by the same token it does not provide any {\it direct} information on the
processes which shaped the Hubble sequence. (3) Although in all our calculations
the IMF extends from 0.1 to 125 $\msun$, by modeling the rest-frame galaxy
luminosity density from 0.15 to 2.2 \micron\ we are actually only sensitive 
to stars within the mass range from $\sim 0.8$ to about 20$\msun$. 
This introduces
non-negligible uncertainties in our estimates of the total amount of stars and
metals produced. (4) No attempt has been made to include the effects of cosmic
chemical evolution on the predicted galaxy colors. All the population synthesis
models assume solar metallicity and will therefore generate colors at
early epochs that are too red.  (5) The uncertanties present in the estimates
of the UV luminosity density from the identification of Lyman-break galaxies in
the HDF are quite large, and the data points at $z>2$ should still be regarded
as tentative. This is especially true for the faint blue dropout sample at
$\langle z\rangle=4$, where  only one spectroscopic confirmation has been 
obtained so far. On the other hand, there is no evidence for a gross
mismatch at the $z\approx 2$ transition between the photometric redshift sample
of Connolly \etal (1997) and the M96 ultraviolet dropout sample. 

\begin{figure}
\plotfiddle{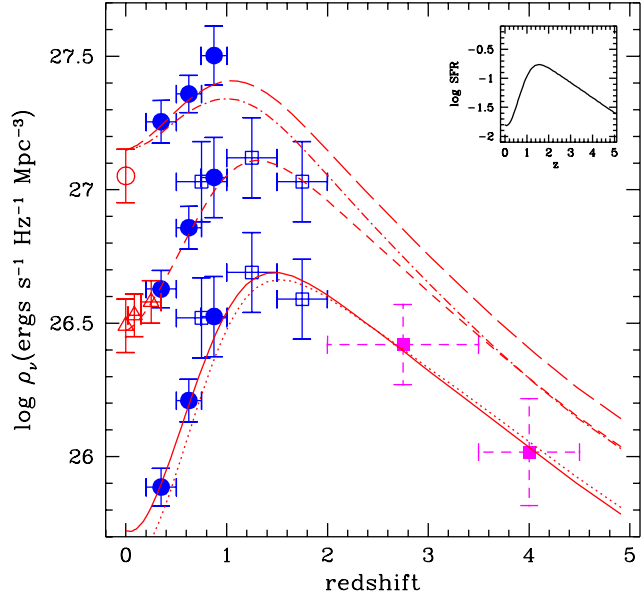}{3.5in}{0}{50}{50}{-150}{-80}
\caption{Same as in Figure 6, but assuming an IMF with $\phi(m)\propto
m^{-1-1.7}$ and no dust extinction. 
}
\label{fig8}
\end{figure}

\subsection{Salpeter IMF}

Figure 7 shows the model predictions for the evolution of $\rho_\nu$ at
rest-frame ultraviolet to near-infrared frequencies.  In the
absence of dust reddening, this relatively flat IMF generates spectra that are
too blue to reproduce the observed mean galaxy colors. The shape of
the predicted and observed $\rho_\nu(z)$ relations agrees better to within the
uncertainties if some amount of dust extinction, $E(B-V)=0.06$, is included.
In this case, the observed UV luminosities must be corrected upwards by a
factor of 1.4 at 2800 \AA\, and 2.1 at 1500 \AA.  As expected, while the
ultraviolet emissivity traces remarkably well the rise, peak, and sharp drop in
the instantaneous star formation rate (the smooth function shown in the inset
on the upper-right corner of the figure), an increasingly large component of
the longer wavelenghts light reflects the past star formation history. The peak
in the luminosity density at 1.0 and 2.2 \micron\ occurs then at later epochs,
while the decline from $z\approx 1$ to $z=0$ is more gentle than observed at
shorter wavelengths. The total stellar mass density at $z=0$ is
$\rho_s(0)=3.7\times 10^8\mdden$, with a fraction close to 65\% being produced
at $z>1$, and only 20\% at $z>2$.  In the assumed
cosmology, about half of the stars observed today are more than 9 Gyr old, and
only 20\% are younger than 5 Gyr.\footnote{Note that, contrary to the measured
number densities of objects and rates of star formation, the integrated stellar
mass density does not depend on the assumed cosmological model.} 

\subsection{x=1.7 IMF}

Figure 8 shows the model predictions for a $x=1.7$ IMF and negligible dust
extinction. While able to reproduce quite well the $B$-band emission history
and consistent within the error with the local $K$-band light, this model
slightly underestimates the 1 \micron\ luminosity density at $z\approx 1$. The
total stellar mass density today is larger than in the previous case,
$\rho_s(0)=6.2\times 10^8\mdden$. 

\subsection{Scalo IMF}

The fit to the data is found to be much poorer for a Scalo function, since 
this IMF
generates spectra that are too red to reproduce the observed mean galaxy
colors, as already noted by Lilly \etal (1996). Because of the relatively large
number of solar mass stars formed, it produces too much long-wavelength light
by the present epoch (Madau \etal 1997). The addition of 
dust reddening would obviously make the fit even worse. 

\subsection{The Brightness of the Night Sky}

Our modeling of the data points to a redshift range $1\lta z\lta 2$  where
the bulk of the stellar mass was actually produced. The uncertainties in the
determination of the luminosity density at that epoch are, however, quite
large. At $z\approx 1$, the increase in the ``estimated'' emissivity 
(i.e., corrected for incompleteness by integrating over the best-fit Schechter
function) over that ``directly'' observed in the CFRS galaxy sample is about 
a factor of 2 (Lilly \etal 1996). Between $z=1$ and $z=2$, the peak in the
average SFR is only constrained by the photometric redshifts of Connolly \etal
(1997) and by the HDF UV dropout sample, both of which may be subject to
systematic biases. 

An important check on the inferred emission history of field galaxies
comes from a comparison of the EBL produced by known galaxies (see \S2.1)
and the predicted mean surface brightness of the night sky, 
\begin{equation}
I_\nu={1\over 4 \pi}\int_0^\infty dz {dl\over dz}\rho_{\nu'}(z)
\end{equation}
where $\nu'=\nu(1+z)$ and $dl/dz$ is the cosmological line element. The results
are plotted in Figure 9. The overall agreement is remarkably good, with the
model spectra being only slightly bluer, by about 20--30\%, than the observed
EBL. The straightforward conclusion of this exercise is that the star formation
histories depicted in Figures 7 and 8 appear able to account for the entire
background light recorded in the galaxy counts down to the very faint
magnitudes probed by the HDF. 

\begin{figure}
\plotfiddle{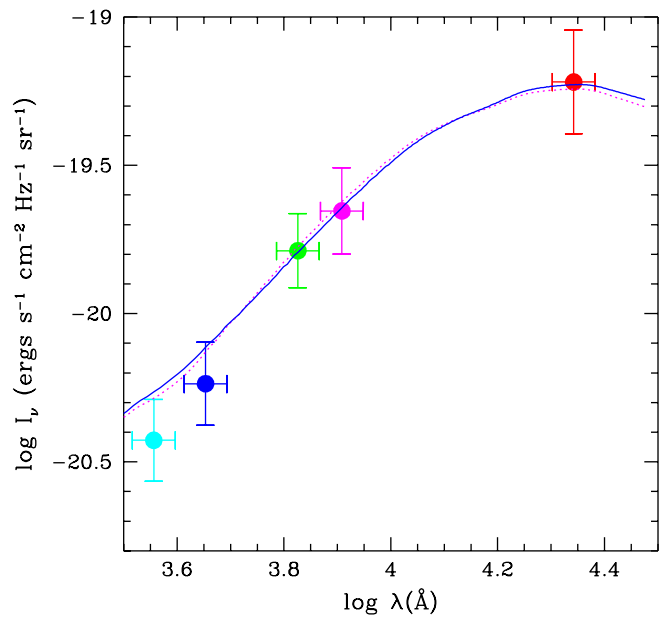}{3.5in}{0}{50}{50}{-150}{-80}
\caption{Spectrum of the extragalactic background light as derived from a 
compilation of ground-based and HDF galaxy counts (see Pozzetti \etal 1997).
The 2$\sigma$ error bars arise mostly from field-to-field variations. {\it
Solid line:} Model predictions for a Salpeter IMF and $E(B-V)=0.06$ (star 
formation history of Figure 7). {\it Dotted line:} Model predictions for 
a $x=1.7$ IMF and negligible dust extinction (star formation history of Figure
8). 
}
\label{fig9} 
\end{figure}

\subsection{The Colors of High-Redshift Galaxies}

Figure 10 shows a comparison between the HDF data and the model
predictions for the evolution of galaxies in the $\ub$ vs. $\vi$
color-color plane according to the star formation histories of Figures 7
and 8. The fact that the Salpeter IMF, $E(B-V)=0.06$ model 
reproduces quite well the rest-frame UV colors of high-$z$ galaxies,
while a dust-free $x=1.7$ IMF generates $\vi$ colors that are 0.2 mag
too blue, suggests the presence of some amount of dust extinction in 
Lyman-break galaxies at $z\sim 3$ (cf. Meurer \etal 1997).

\begin{figure}
\plotfiddle{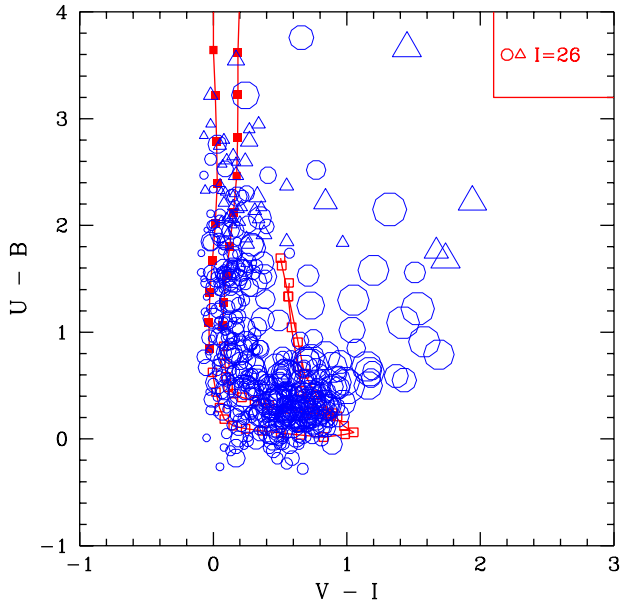}{3.5in}{0}{50}{50}{-150}{-80}
\caption{{\it Solid lines:} model predictions for the color evolution of 
galaxies according to the star formation histories of Figure 7 ({\it
right curve}) and 8 ({\it left curve}). The points 
({\it filled squares} for $z>2$ 
and {\it empty squares} for $z<2$) are plotted at redshift interval $\Delta 
z=0.1$. {\it Empty circles} and {\it triangles:} colors of galaxies in the HDF 
with $22<B<27$. The symbol types and sizes are as described for Figure 5. 
The ``plume'' of reddened high-$z$ galaxies is clearly seen in the data. 
}
\label{fig10} 
\end{figure}

\subsection{The Stellar Mass Density Today}

The best-fit models discussed in \S\,4 generate a present-day stellar mass
density in the range between 4 and 6 $\times 10^8\mdden$. Although one could in
principle lower this number by adopting a top-heavy IMF, richer in massive
UV-producing stars, in practice a significant amount of dust reddening -- hence
of ``hidden'' star formation -- would then be required to match the observed
galaxy colors. The net effect of this operation would be a baryonic mass
comparable to the estimate above and a far too large infrared background (see
below). It appears therefore that the observed galaxy emission history implies
a stellar mass density at the present epoch in the interval $0.005\lta
\Omega_sh_{50}^2\lta 0.009$. The stellar mass-to-light ratios range from 4.5 in
the $B$-band and 0.9 in $K$ for a Salpeter function, to 8.1 in $B$ and 1.5 in
$K$ for a $x=1.7$ IMF. Note that these values are quite sensitive to the
lower-mass cutoff of the IMF, as very-low mass stars can contribute
significantly to the mass but not to the integrated light of the whole stellar
population. A lower cutoff of 0.2$\msun$, instead of the 0.1$\msun$ adopted,
would decrease the mass-to-light ratio by a factor of 1.3 for a Salpeter
function, 1.6 for $x=1.7$, and 1.1 for a Scalo IMF.

\section{Star Formation at High Redshift: Monolithic Collapse Versus 
Hierarchical Clustering Models} 

The significant uncertanties present in the estimates of the star formation 
density at $z>2$ have already been mentioned. The biggest  one is
probably associated with dust reddening, but, as the color-selected HDF sample
includes only the most actively star-forming young objects, one could also
imagine the existence of a large population of relatively old or faint galaxies
still undetected at high-$z$. The issue of the amount of star formation at
early epochs is a non trivial one, as the two competing models, ``monolithic
collapse'' versus hierarchical clustering, make very different predictions in
this regard. From stellar population studies we know in fact that about half of
the present-day stars are contained into spheroidal systems, i.e., elliptical
galaxies and spiral galaxy bulges (Schechter \& Dressler 1987). In the
``monolithic'' scenario these formed early and rapidly, experiencing a bright
starburst phase at high-$z$ (Eggen, Lynden-Bell, \& Sandage 1962; Tinsley \&
Gunn 1976; Bower \etal 1992). In hierarchical clustering theory instead
ellipticals form continuosly by the merger of disk/bulge systems (Kauffman
\etal 1993; White \& Frenk 1991), and most galaxies never experience star 
formation rates in excess
of a few solar masses per year (Baugh \etal 1997). The star formation
histories discussed in \S~4 produce only 10\% of the current stellar content of
galaxies at $z>2.5$, in apparent agreement with hierarchical clustering
cosmologies. In fact, the tendency to form the bulk of the stars at relatively
low redshifts is a generic feature not only of the $\Omega_0=1$ CDM cosmology,
but also of successful low-density CDM models (cf Figure 21 of Cole \etal 
1994; Baugh \etal 1997).

It is then of interest to ask how much larger could the volume-averaged SFR at
high-$z$ be before its fossil records -- in the form of long-lived, near
solar-mass stars -- became easily detectable as an excess of $K$-band light at
late epochs. In particular, is it possible to envisage a toy model where 50\%
of the present-day stars formed at $z>2.5$ and were shrouded by dust? The
predicted emission history from such a model is depicted in Figure 11. To
minimize the long-wavelength emissivity associated with the radiated
ultraviolet light, a Salpeter IMF has been adopted. Consistency with the HDF
data has been obtained assuming a dust extinction which increases rapidly with
redshift, $E(B-V)=0.0067(1+z)^{2.2}$. This results in a correction to the rate
of star formation of a factor $\sim 5$ at $z=3$ and $\sim 15$ at $z=4$. 
The total stellar mass density today is $\rho_s(0)=5.0\times 10^8\mdden$ ($\Omega_sh_{50}^2=$0.007). 

Overall, the fit to the data is still acceptable, showing how the blue and
near-IR light at $z<1$ are relatively insensitive to significant variations in
the SFR at high redshifts, and are then, because of the short 
timescale available at $z\gta 2$, {\it  relatively poor indicators of the star
formation history at early epochs}. Note, however, that the adopted
extinction-redshift relation implies negligible reddening at $z\lta 1$.
Relaxing this -- likely unphysical -- assumption would cause the model to
significantly overproduce the $K$-band local luminosity density. We have also
checked that a larger amount of hidden star formation at early epochs would
generate too much blue, 1 \micron\, and 2.2 \micron\ light to be still
consistent with the observations. An IMF which is less rich in massive stars
would only exacerbate the discrepancy. 

\begin{figure}
\plotfiddle{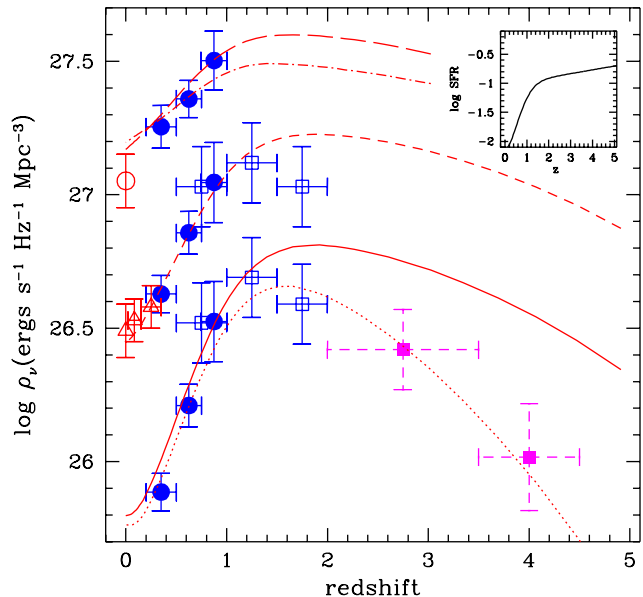}{3.5in}{0}{50}{50}{-150}{-80}
\caption{Test case with a much larger star formation density at high redshift
than indicated by the HDF dropout analysis. The model -- designed to mimick a
``monolithic collapse'' scenario -- assumes a Salpeter IMF and a dust opacity
which increases rapidly with redshift, $E(B-V)=0.0067(1+z)^{2.2}$. Notation is
the same as in Figure 7. 
\label{fig11}}
\end{figure}

\subsection{Constraints from the Mid- and Far-Infrared Background}

Ultimately, it should be possible to set some constraints on the total amount
of star formation hidden by dust over the entire history of the universe
by looking at the cosmic infrared background (CIB). From an analysis of the
smoothness of the {\it COBE} DIRBE maps, Kashlinsky \etal (1996) have recently
set an upper limit to the CIB of 10--15 nW m$^{-2}$ sr$^{-1}$ at
$\lambda=$10--100 \micron\ assuming clustered sources which evolve according to
typical scenarios. An analysis using data from {\it COBE} FIRAS by Puget \etal
(1996) has produced a tentative detection at a level of 3.4 ($\lambda/400
\micron)^{-3}$ nW m$^{-2}$ sr$^{-1}$ in the 400--1000 \micron\ range.  By
comparison, the integrated light that is reprocessed by dust in the model
depicted in Figure 7 is close to 8 nW m$^{-2}$ sr$^{-1}$. The monolithic
collapse scenario of Figure 11 generates about 6.5 nW m$^{-2}$ sr$^{-1}$ 
instead.  While both these models appear to be consistent with the data 
(given the large 
uncertainties associated with the subtraction of foreground emission and 
the spectral shape of the CIB), it is clear that a significantly larger amount
of hidden star formation at early and/or late epochs would overproduce
the IR background (Fall \etal 1996; Guiderdoni \etal 1997).

\subsection{Metal Production}

We may at this stage use our set of models to establish a cosmic
timetable for the production of heavy elements ($Z\ge 6$) in relatively bright
field galaxies. What we are interested in here is the universal rate
of ejection of newly synthesized material. In the approximation of
instantaneous recycling, the metal ejection rate per unit comoving volume can
be written as 
\begin{equation}
{\dot \rho_Z}=y(1-R)\times {\rm SFR}, \label{eq:rhoz}
\end{equation}
where the {\it net}, IMF-averaged yield of returned metals is 
\begin{equation}
y={\int mp_{\rm zm}\phi(m)dm\over (1-R)\int m\phi(m)dm},
\end{equation}
$p_{\rm zm}$ is the stellar yield, i.e., the mass fraction of a star of
mass $m$ that is converted to metals and ejected, and the dot denotes
differentiation with respect to cosmic time. 

\begin{figure}
\plotfiddle{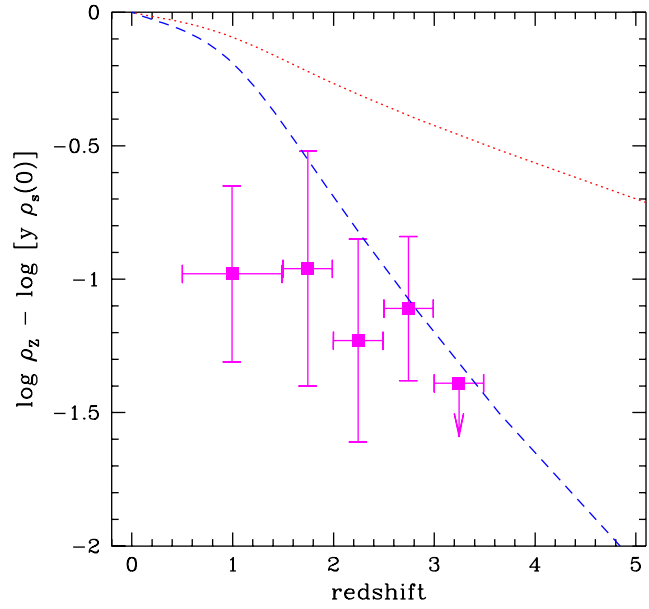}{3.5in}{0}{50}{50}{-150}{-80}
\caption{Total mass of heavy elements ever ejected versus redshift for the
Salpeter IMF model of Figure 7 ({\it dashed line}) and the ``monolithic
collapse'' model of Figure 11 ({\it dotted line}), normalized to $y\rho_s(0)$,
the total  mass density of metals at the present epoch. {\it Filled squares:}
column density-weighted metallicities (in units of solar) as derived from
observations of the damped Lyman-$\alpha$ systems (Pettini \etal 1997). 
\label{fig12}}
\end{figure}

The predicted end-products of stellar evolution, particularly from massive
stars, are subject to significant uncertainties. These are mainly due to the
effects of initial chemical composition, mass-loss history, the mechanisms of
supernova explosions, and the critical mass, $M_{\rm BH}$, above which stars
collapse to black holes without ejecting heavy elements into space
(Maeder 1992; Woosley \& Weaver 1995). The IMF-averaged yield 
is also very sensitive to the choice of the IMF slope and lower-mass cutoff.
Observationally, the best-fit ``effective yield'' (derived assuming a closed
box model) is 0.025$Z_\odot$ for Galactic halo clusters, 0.3$Z_\odot$ for disk
clusters, 0.4$Z_\odot$ for the solar neighborhood, and 1.8$Z_\odot$ for the
Galactic bulge (Pagel 1987). The latter value may represent the universal true
yield, while the lower effective yields found in the other cases may be due,
e.g., to the loss of enriched material in galactic winds. 

Figure 12 shows the total mass of metals ever ejected, $\rho_Z$, versus 
redshift, i.e., the sum of the heavy elements stored in stars and in the gas
phase as given by the integral of equation (\ref{eq:rhoz}) over cosmic time.
The values plotted have been computed from the star formation histores 
depicted in Figures 7 and 11, and have been normalized to $y\rho_s(0)$,
the mass density of metals at the present epoch according to
each model. A characteristic feature of the two competing scenarios is the
rather different average metallicity expected at high redshift. For comparison,
we have also plotted the {\it gas metallicity}, $Z_{\rm DLA}/Z_\odot$, as
deduced from observations by Pettini \etal (1997) of the damped Lyman-$\alpha$
systems (DLAs).  At early epochs,  when the gas consumption into stars is still
low, the metal mass density predicted from these models gives, in a closed box
model, a measurement of the metallicity of the gas phase. If DLAs and
star-forming field galaxies have the same level of heavy element enrichment
(Pei \& Fall 1995),
then one would expect a rough agreement between $Z_{\rm DLA}$ and the model
predictions at $z\gta 3$. This is not true at $z\lta 2$, when a significant
fraction of heavy elements is locked into stars.  Without reading too much into
this comparison (note the large error bars associated with the data points), 
it does appear
that the monolithic collapse model tends to overpredict the cosmic metallicity
at high redshifts as sampled by the DLAs. 

\bigskip

While the detection with NICMOS of the established stellar populations
surrounding the regions of star formation observed in the HDF at $z\sim
2$ should shed some light on the questions addressed in this talk, it will
ultimately take the Next Generation Space Telescope to see the visible light
emitted by stars at $z=2-5$, and to effectively open much of the
universe to a direct study of galaxy formation and the history of the 
conversion of neutral gas into stars.

\acknowledgments

I would like to thank G. Bruzual, S. Charlot, A. Connolly, M. Pettini,
and my collaborators, L. Pozzetti and M. Dickinson, for many stimulating 
discussion on various topics related to this talk. Support for this
work was provided by NASA through grant AR-06337.10-94A from the
Space Telescope Science Institute, which is operated by the Association of
Universities for Research in Astronomy, Inc., under NASA contract NAS5-26555.

\end{document}